\documentclass[onecollarge,natbib,sort&compress]{svjour2}
\bibpunct{[}{]}{;}{n}{}{,} 
\smartqed  
\journalname{Archive of Applied Mechanics}

\usepackage{amssymb}
\usepackage{amsmath}
\usepackage{graphicx}
\usepackage{dcolumn}
\usepackage{bm}
\usepackage{epsfig}
\usepackage{amssymb}
\usepackage{color}
\usepackage{datetime}
\usepackage{wasysym}
\usepackage{slashed}
\usepackage[mathscr]{euscript}

\newcommand{\bal}{\begin{align}}
\newcommand{\eal}{\end{align}}
\newcommand{\beq}{\begin{eqnarray}}
\newcommand{\eeq}{\end{eqnarray}}
\newcommand{\nneeq}{\nonumber \end{eqnarray}}

\newcommand{\nn}{\nonumber \\}
\newcommand{\es}{& = &}
\newcommand{\rs}{\, = \,}
\newcommand{\ps}{& + &}

\newcommand{\np}{\nn \ps}

\newcommand{\cA}{ {\cal A} }
\newcommand{\cB}{ {\cal B} }

\newcommand{\cC}{ {\cal C} }
\newcommand{\cD}{ {\cal D} }
\newcommand{\cM}{ {\cal M} }
\newcommand{\cH}{ {\cal H} }

\newcommand{\cG}{ {\cal G} }

\newcommand{\cU}{ {\cal U} }
\newcommand{\cL}{ {\cal L} }

\newcommand{\h}{ {1 \over 2} }

\begin{document}

\title{From asymptotic freedom toward heavy quarkonia \\
within the renormalization group procedure for effective particles\thanks{Invited talk given at Light Cone 2016, IST, Universidade de Lisboa, Portugal, September 5-8 2016}
}


\author{Mar\'ia G\'omez-Rocha 
}


\institute{Mar\'ia G\'omez-Rocha \at European Centre for Theoretical Studies in Nuclear Physics and Related Areas (ECT*), Strada delle Tabarelle, 286  I-38123 Villazzano (TN) \\
              \email{gomezr@ectstar.eu}           
}

\date{Received: date / Accepted: date}

\maketitle

\begin{abstract}
The renormalization group procedure for effective particles (RGPEP), developed as a nonperturbative tool for constructing bound states in quantum field theories, is applied to QCD.
The approach stems from the similarity renormalization group and introduces the concept of effective particles. 
It has been shown that the RGPEP passes the test of exhibiting asymptotic freedom. We present the running of the Hamiltonian coupling with the renormalization-group scale and summarize the basic elements needed in the formulation of the bound-state problem.
\end{abstract}

\section{Introduction}
\label{intro}

The renormalization group procedure for effective particles (RGPEP) has been developed during the last years as a non perturbative tool for constructing bound states in quantum field theory (QFT)~\cite{GlazekCondensatesAPP, GlazekPerturbative, TrawinskiConfinement, GlazekQED, AF-GomezRocha-Glazek}. It stems from the similarity renormalization group~\cite{GlazekWisonHamiltonians,Wilson:1994fk} and introduces the concept of effective particles in the Fock space~\citep{GlazekLC2016}. The RGPEP is a Hamiltonian approach and it is formulated in the front form (FF) of relativistic dynamics~\citep{Dirac}. The FF possesses several interesting features. For example, in the FF Hamiltonian there are no terms with only creation or only annihilation operators;  then, the vacuum state $| 0 \rangle$ is an eigenstate of the Hamiltonian with eigenvalue 0. Furthermore, from the 10 generators of the Poincar\'e group, only 3 are dynamical (contain interactions). This is interesting e.g. from the point of view of situations in which one needs to consider bound states in different reference frames. 

One of the central problems in describing bound states in QFT is the fact that one needs to deal with an infinite number of degrees of freedom. For example, a quarkonium state in the Fock state can be expressed by an infinite series with the structure
\begin{eqnarray}
|\psi\rangle \es |Q\bar Q\rangle + |Q\bar Qg\rangle + |Q\bar Q gg \rangle + \dots \ ,
\label{QbarQ}
\end{eqnarray}
and there is no limit in the number of participating Fock components.  
The key idea of the RGPEP is that one can derive an effective Hamiltonian $H_s$, written in a scale-dependent basis such that, for a certain scale $s$, the number of relevant Fock components of its corresponding eigenstates (cf. Eq.~(\ref{QbarQ})) is small. The parameter $s$ has the meaning of size of the effective particles. When infinitely many Fock components can be neglected and only a few are significant, the bound state problem is drastically simplified and one can attempt to seek for a numerical solution to the eigenvalue equation. We present the main steps of this method in Sections~\ref{SecInitial} and~\ref{SecRGPEP}.

The notion of effective particles is also used to explain the different
behavior of interacting particles at different energy scales. It has been checked that the RGPEP passes the test of asymptotic freedom~\cite{AF-GomezRocha-Glazek}. It will be shown below how the strength of the running coupling depends on the size of the effective particles. 
Exhibiting asymptotic freedom is a precondition for any approach aiming at describing hadrons using QCD. This issue is introduced in Section~\ref{SecAF}.

In order to start thinking about bound states in QCD we focus on heavy quarkonia. 
Heavy quarkonium is the simplest bound state in QCD  
and it provides a fruitful ground
for comparing QCD approaches with experimental data~\cite{Brambilla,Pineda,Hilger:2014nma, Hilger:2015hka,Gomez-Rocha:2015qga,Gomez-Rocha:2016cji, BlankBottomonium,Popovici,Fischer:2014cfa, LiQuarkonium,Segovia,Greensite,GuoSzczepaniakHybrids, Mlynik}. A number of effective theories and
phenomenological models are based on simplifications appearing due to the very large quark masses as compared with $\Lambda_{QCD}$. 
 This fact and the asymptotic freedom feature will allow us to construct an effective theory using RGPEP. We present the main steps and basic elements of the bound-state problem for heavy quarkonia in Section~\ref{SecQbarQ}.

\section{The initial Hamiltonian} 
\label{SecInitial}

\subsection{The QCD canonical Hamiltonian} 

We start from the classical Lagrangian of QCD:
\begin{eqnarray}
\cL\es \bar \psi (i\slashed D - m)\psi - \h \text{tr} F^{\mu\nu}F_{\mu\nu} \ ,
\end{eqnarray}
where $F^{\mu\nu} = {1 \over ig} [D^\mu,D^\nu]$, $D = \partial + ig A $, $A = A^a t^a$ and $[t^a,t^b] = i f^{abc} t^c$ .
We use the front form of dynamics and light-front variables (we adopt the conventions of Ref.~\citep{Brodsky-Pauli-Pinsky}). The Noether theorem provides the associated energy-momentum tensor which, in the gauge $A^+=0$ and with the choice of the initial conditions set on the hypersurface $x^+=x^0+x^1=0$, leads to the following QCD front-form Hamiltonian
\begin{eqnarray}
H_{QCD} \ = \ P^- \es {1 \over 2}\int dx^- d^2 x^\perp {\cal H}\, |_{x^+=0} \quad ,
\label{Pm}
\end{eqnarray}
where
\begin{eqnarray}
\cH = \cH_{\psi^2} + \cH_{A^2} + \cH_{A^3} + \cH_{A^4} + \cH_{\psi A \psi} + 
\cH_{\psi A A \psi} + \cH_{[\partial A A]^2} + \cH_{[\partial A A](\psi\psi)} 
+ \cH_{(\psi\psi)^2} \ ,
\label{Hdensity}
\end{eqnarray}
with the expression for every term given in the Appendix~\ref{AppCanonical}. 

At the quantization surface $x^+=0$, the quantum fields are defined by\footnote{In the sequel we will drop the hats in particle operators, for simplicity.}
\begin{eqnarray}
\hat A^\mu \es \sum_{\sigma c} \int [k] \left[ t^c \varepsilon^\mu_{k\sigma}
\hat a_{k\sigma c} e^{-ikx} + t^c \varepsilon^{\mu *}_{k\sigma}
\hat a^\dagger_{k\sigma c} e^{ikx}\right]_{x^+=0} \ , \\
\label{Amu}
\hat\psi \es \sum_{\sigma c f} \int [k] 
  \left[ \chi_c u_{fk\sigma} \hat b_{k\sigma c f} e^{-ikx} + 
  \chi_c v_{fk\sigma} \hat d^\dagger_{k\sigma c f} e^{ikx}
  \right]_{x^+=0} \ ,
\end{eqnarray}
with the shorthand notation for the integration measure $[k] = \theta(k^+)
dk^+ d^2 k^\perp/(16\pi^3 k^+) $, the gluon polarization four-vector $\varepsilon^\mu_{k\sigma} 
= (\varepsilon^+_{k\sigma}=0, \varepsilon^-_{k\sigma} 
= 2k^\perp \varepsilon^\perp_\sigma/k^+, 
\varepsilon^\perp_\sigma)$,  and $u_{fk\sigma}$ and $v_{fk\sigma}$ being Dirac spinors. The indices $\sigma$ and $c$ stand for spin polarization and color, respectively. The replacement in Eq.~(\ref{Hdensity}) yields the canonical QCD Hamiltonian, a quantum operator written in terms of creation and annihilation operators (cf. Refs.~\citep{GlazekHO,Mlynik} for details in these expressions). 

\subsection{Regularization}
The canonical Hamiltonian needs regularization. We will follow the same procedure as in Ref.~\citep{Glazek2000}.
Every interaction term in the Hamiltonian is labeled by its momentum quantum numbers $p^\perp$ and $p^+$; the total momentum annihilated in a vertex is labeled by $P^\perp$ and $P^+$; the relative transverse momentum is given by $\kappa^\perp = p^\perp - x P^\perp$, and the longitudinal momentum fraction by $x=p^+/P^+$.
In every term in the Hamiltonian, every creation and annihilation operator of any momentum $p$ is multiplied by the regulating function
\begin{eqnarray}
r_{\Delta \delta}(\kappa^\perp, x) 
\es 
\exp{(- \kappa^{\perp 2}/\Delta^2)} \, x^\delta \, \theta(x) \ .
\end{eqnarray}
The first factor regulates ultraviolet divergences appearing at large $\kappa^{\perp 2}$. The Gaussian function does not allow the change of any gluon relative transverse momentum $\kappa^\perp$ to exceed the large cutoff $\Delta$. The second factor, depending on the small cutoff parameter $\delta$, regulates divergences appearing due to small-$x$ in denominators. 

After the derivation of the effective Hamiltonian (see below), any remaining cutoff-dependent term in the limit $\Delta\to\infty$ and $\delta\to 0$ must be canceled by counterterms. 
The regularized canonical Hamiltonian plus counterterms, provide the initial condition for the RGPEP equation.  

\section{Renormalization group and effective particles}
\label{SecRGPEP}
The RGPEP introduces the concept of \textit{effective particles}~\cite{GlazekLC2016}. The initial Hamiltonian $H_0$, which is written in term of (bare) creation and annihilation operators of pointlike particles (quark, antiquark or gluons) $q_0^\dagger$ and $q_0$ can be re-expressed in terms of particle operators of size $s$. Creation (annihilation) operators of effective particles labeled with $s$ acting on the Fock space create (annihilate) effective particles of size $s$. E.g. for gluons:
\begin{eqnarray}
\quad a_0^\dagger |0\rangle =  |g\rangle \ ,  \quad  
\quad a_s^\dagger |0\rangle =  |g_s\rangle \ , \quad
a_0|0\rangle =  a_s|0\rangle = 0\ .
\end{eqnarray}
It is also common to use the momentum-width parameter $\lambda=1/s$. This parameter distinguishes different kinds of gluons according to the rule that: effective gluons of type $\lambda$ can change their relative motion kinetic energy though a single effective interaction by no more than about $\lambda$~\citep{Glazek2000}. It is however more convenient now to illustrate the calculation using the parameter $t=s^4=1/\lambda^4$.  

In terms of this parameter, bare and effective particle operators are related by the unitary transformation: 
\begin{eqnarray}
q_t \es \cU_t \, q_0 \, \cU_t^\dagger \ , 
\qquad {\rm with} \qquad
\cU_t
\ = \ 
T \exp{ \left( - \int_0^t d\tau \, \cG_t
\right) } \ , \label{Ut}
\label{aU}
\end{eqnarray}
where $\cG_t$ is a generator that will be given later. 
The RGPEP is based on the condition that the effective Hamiltonian cannot depend on the scale parameter $t$:
\begin{eqnarray}
H_t(q_t) \es H_0 (q_0) \ . \label{condition}
\end{eqnarray}
The relation between $q_t$ and $q_0$ implies that 
\begin{eqnarray}
H_t(q_0) = \cU_t^\dagger H_0 (q_0)\cU_s \equiv \cH_t \ .
\end{eqnarray}
The differentiation of this equation leads to the RGPEP equation:
\begin{eqnarray}
\cH_t' 
\es \left[ \cG_t ,
\cH_t \right] \ .
\label{RGPEP}
\end{eqnarray}
We consider the generator $\cG_t=[ \cH_f, \cH_{Pt} ]$ (see Appendix~\ref{AppRGPEP} for details on the notation). Other generators are also possible, like the one used in Ref.~\citep{Glazek2000}, but they may lead to more complicated expressions. 

The initial condition needed for solving Eq.~(\ref{RGPEP}) is given by the initial Hamiltonian at $t=0$ plus counterterms:
\begin{eqnarray}
\cH_{t=0} \es \cH_{\Delta\delta}  + {\rm CT}_{\Delta\delta} \ .
\end{eqnarray}
Although in principle Eq.~(\ref{RGPEP}) can be solved nonperturbatively, we are not prepared yet to do so in QCD. We use the perturbative formulas presented in Ref.~\citep{GlazekPerturbative} and solve it order by order as given in the Appendix~\ref{AppRGPEP}.

\section{Asymptotic freedom}
\label{SecAF}

Already at third order, the effective Hamiltonian $H_t$ contains details of the structure of the three-gluon vertex and quark-gluon vertex. We have studied the three-gluon vertex in Ref.~\citep{AF-GomezRocha-Glazek} and calculated the Hamiltonian running coupling. The result,
\beq
\label{gl}
g_\lambda \es
g_0 - { g_0^3 \over 48 \pi^2 }   N_c \,   11 \,\ln
{ \lambda \over \lambda_0} \ ,
\eeq 
leads to the $\beta$-function that is familiar to us and is found in Refs.~\citep{Gross:1973id,Politzer:1973fx} if one identifies $\lambda$ with the momentum scale of external gluon lines in Feynman diagrams,
\beq
\lambda {d \over d\lambda} \, g_\lambda
\es \beta_0 g_\lambda^3 \ , \qquad {\rm with} \qquad \beta_0 \rs - { 11 N_c \over 48\pi^2 } \ .
\eeq

\begin{figure}[h]
\centering
\sidecaption
\includegraphics[scale=0.7]{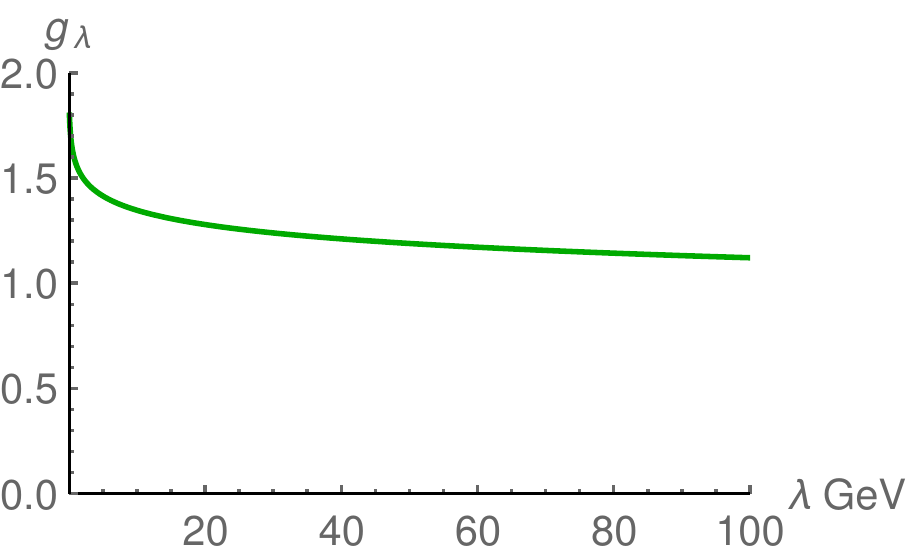}\hspace{0.5cm}
\includegraphics[scale=0.7]{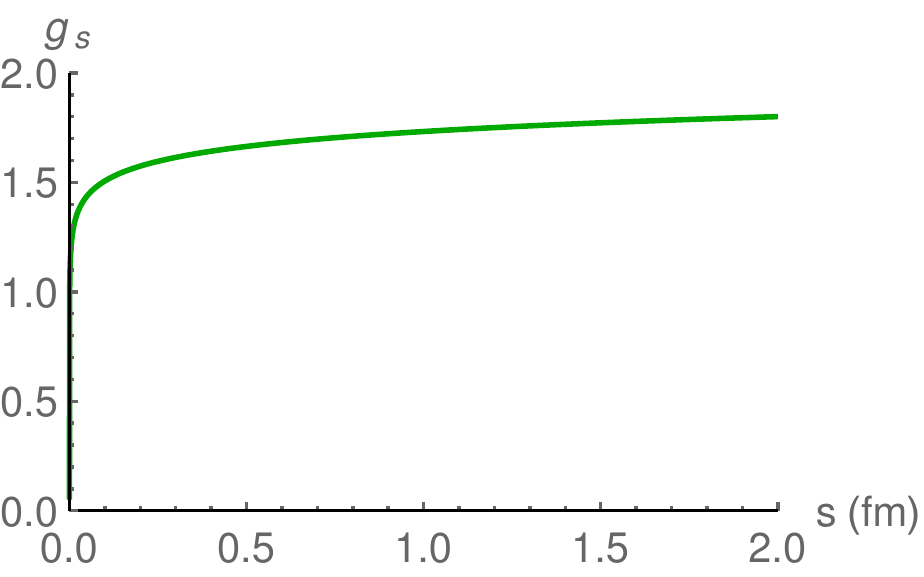}
\caption{Running coupling as function of the momentum-width parameter $\lambda$ (left) and as a function of the size of effective particles $s$ (right).}
\label{Fig_g}
\end{figure}

The dependence on the running coupling with $\lambda$ and $s$ is plotted in Fig.~\ref{Fig_g}. The smaller is the size of the effective particles, the weaker is the interaction between them; this corresponds to the situation of asymptotic freedom. This situation is explained in a visual way using a picture in Ref.~\citep{GlazekCondensatesAPP}. 

This result suggests a possible direction in which one could start thinking about the equivalence of the
Euclidean Green’s function calculus and Minkowskian Hamiltonian quantum field operator.

\section{Toward heavy quarkonium}
\label{SecQbarQ}

\subsection{A heavy-quark effective theory}

\begin{figure}[h]
\centering
\sidecaption
\includegraphics[scale=0.5]{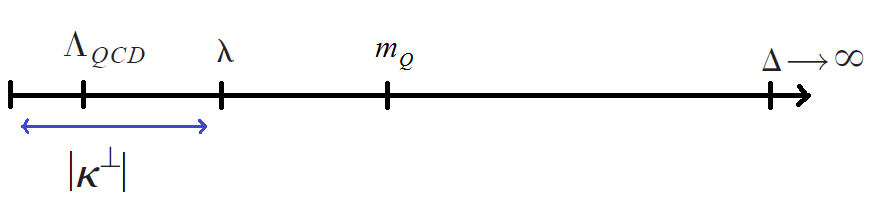}
\caption{Hierarchy of scales in the heavy-quark effective theory. Effective particles cannot change their relative momentum $|\kappa^\perp|$ by more than about $\lambda$. In turn, this parameter is much smaller than the quark mass $m_Q$. The ultraviolet parameter $\Delta$ is much larger than any other momentum parameter and is sent to infinity.}\label{FigHierarchy}
\end{figure}

The RGPEP allows one for the construction of effective theories. We will consider a theory of very heavy quarks, with masses much larger than the scale parameter $\lambda$ (so that $\lambda/m_Q\to 0$), and with only one flavor in the QCD Hamiltonian. In this simplified situation, the coupling constant can still be considered small by means of asymptotic freedom and one can use expansions in powers on $g$ when solving the RGPEP equation. Our hierarchy of scales is sketched in Fig.~\ref{FigHierarchy}.

\subsection{The bound-state problem}
In our heavy quark effective theory, in which only one (heavy) flavor is present, it is not possible to create quark-antiquark pairs of such large masses out of an intermediate gluon with momentum not larger than $\lambda$. Hence, the higher Fock component following to the simple $|Q\bar Q\rangle$ sector contain only additional gluons. The minimum number of sectors necessary to account for non-Abelian contributions is three:
\begin{eqnarray}
| \psi \rangle
=
| Q\bar Q \rangle 
+
| Q\bar Q \, g\rangle 
+
| Q\bar Q \, gg\rangle  \ . 
\end{eqnarray}
Talking about three Fock components in our perturbative RGPEP formalism is the same thing as talking about 4th-order calculations. 

We follow Ref.~\citep{WilsonBoundStateH} and use projection operators to separate Fock components and formulate the eigenvalue equation in terms of the $|Q\bar Q\rangle$ state. As distinct from~\citep{WilsonBoundStateH}, here the operators $\hat P_t$ and $\hat Q_t=(1-\hat P_t)$ depend on $t$ and act on Fock sectors of particles with size $s=t^{1/4}$ :
\begin{eqnarray}
\psi_P \equiv  \hat P_t | \psi \rangle  \es | Q\bar Q \rangle \ ,    \\
\psi_Q \equiv \hat Q_t | \psi \rangle \es| Q\bar Q \, g\rangle  
+
| Q\bar Q \, g\rangle   \ ,
\end{eqnarray}
with
$ \psi  =  \psi_P  +  \psi_Q $. 
States $\psi_P$ and $\psi_Q$ are related by the rotation\footnote{We drop hats again in the operators from now on.}
$\psi_Q  = \hat R_t \, \psi_P$. 
The eigenvalue equation reads
\begin{eqnarray}
H_t(\psi_P+\psi_Q)=E(\psi_P+\psi_Q) \ .
\end{eqnarray} 
Projecting the eigenvalue equation once into the subspace $P$ and once into $Q$ and using $\hat R_t$, one can obtain after careful manipulations~\cite{WilsonBoundStateH} the effective Hamiltonian for $| Q\bar Q \rangle$ :
\begin{eqnarray}
 H_{{\rm eff} \, t} | Q\bar Q \rangle  = E | Q\bar Q \rangle  \ ,
\end{eqnarray}
where
\beq
 H_{{\rm eff} \, t} \es{1\over \sqrt{ P + R_t^\dagger R_t} } (P_t+R_t^\dagger)H_{t} (P_t+R_t){1\over \sqrt{ P + R_t^\dagger R_t} } \ . 
\eeq
When one assumes the coupling be small, like it is  in our heavy-quark effective theory, one can use expansions in powers of $g$ also to derive $R_t$ and last $ H_{{\rm eff} \, t}$. It should be emphasized that perturbation theory is only employed to solve the RGPEP equation and to calculate $R_t$, but not in solving the eigenvalue problem. 

Second-order calculations using a particular generator (different from ours) have been performed, and the spectrum of a large number of quarkonia~\citep{GlazekHO,Mlynik} has been obtained. However, details that can provide new hints about deeper aspects of the theory, such as e.g. relevant information related to the confinement mechanism, can appear only starting from fourth order. 

Another important feature of the RGPEP is that, since it uses a formulation in the Fock space which includes gluon degrees of freedom explicitly, the method provides an ideal framework for the description of hybrids~\cite{HybridsSZ,HybridsJ}.

\section{Summary}

We have presented the RGPEP as a tool to deal with bound states in QCD using the concept of effective particles. We have given the perturbative formula that solves the RGPEP equation and provides third- and fourth-order renormalized Hamiltonians. 

The third-order $H_t$ contains the running of the Hamiltonian coupling constant. We have presented the obtained result and its variation with the scale parameter. The derived running coupling has the same asymptotic-freedom behavior as in the calculus based
on renormalized Feynman diagrams. This is important from the point of view of the desired but
unknown precise connection between Feynman diagrams for virtual transition amplitudes and the
Hamiltonan formalism in the Minkowski spacetime.

Heavy-quarkonium is the simplest bound state in QCD and can be studied in the context of an effective theory for heavy quarks within the RGPEP. In order to investigate non-Abelian contributions and thus to include explicitly the running of the coupling in the bound-state problem, at least fourth-order effective Hamiltonians are necessary. Fourth-order calculations are drastically more complicated than third-order ones.

\begin{acknowledgements}
I am very grateful to the Gary McCartor Fellowship Program for the McCartor travel award, thanks to which I could attend to this conference. I congratulate the LC2016 organizers for the very interesting and enjoyable  meeting. 
\end{acknowledgements}


\begin{appendix}

\section{Hamiltonian-density terms}
\label{AppCanonical}
The terms in Eq.~(\ref{Hdensity}) are 
\beq
 {\cal H}_{\psi^2} \es {1\over 2} \bar \psi \gamma^+ {-\partial^{\perp \, 2} + 
  m^2 \over i\partial^+} \psi  \ ,
\\
{\cal H}_{A^2} \es - {1\over 2} A^{\perp a } (\partial^\perp)^2 A^{\perp a} \  , 
\\
{\cal H}_{A^3} \es   g \, i\partial_\alpha A_\beta^a [A^\alpha,A^\beta]^a  \ ,
\\
{\cal H}_{A^4} \es  - {1\over 4} g^2 \, [A_\alpha,A_\beta]^a[A^\alpha,A^\beta]^a  \ , 
\\
{\cal H}_{\psi A \psi} \es g \, \bar \psi \hspace{-4pt}\not\!\!A \psi  \ ,
\\
 {\cal H}_{\psi A A \psi} \es  {1\over 2}g^2 \, \bar \psi \hspace{-4pt}\not\!\!A 
  {\gamma^+ \over i\partial^+} \hspace{-4pt}\not\!\!A \psi \ ,
\\
  {\cal H}_{[\partial A A]^2} \es {1\over 2}g^2 \, 
  [i\partial^+A^\perp,A^\perp]^a {1 \over (i\partial^+)^2 }
  [i\partial^+A^\perp,A^\perp]^a \ ,
\\
  {\cal H}_{[\partial A A](\psi\psi)} \es g^2 \, 
  \bar \psi \gamma^+ t^a \psi {1 \over (i\partial^+)^2 }
  [i\partial^+A^\perp,A^\perp]^a, 
\\
  {\cal H}_{(\psi\psi)^2} \es {1\over 2}g^2 \,
  \bar \psi \gamma^+ t^a \psi {1 \over (i\partial^+)^2 }
  \bar \psi \gamma^+ t^a \psi \ .
\eeq

\section{Summary of RGPEP formulas}
\label{AppRGPEP}

The RGPEP equation satisfied by the renormalized Hamiltonian $\cH_t \equiv H_t(q_0) = \cU_t^\dagger H_0 (q_0)\cU_s$ is:
\begin{eqnarray}
\cH_t' &=& [[\cH_f,\cH_{Pt}],\cH_t]
\end{eqnarray}
where
\begin{eqnarray}
\cH_f \es
\sum_i \, p_i^- \, q^\dagger_{0i} q_{0i} \ , \qquad {\rm with} \qquad  p^-_i = { p_i^{\perp \, 2} \over p_i^+} \ , \\
\cH_t(q_0) \es
\sum_{n=2}^\infty \, 
\sum_{i_1, i_2, ..., i_n} \, c_t(i_1,...,i_n) \, \, q^\dagger_{0i_1}
\cdot \cdot \cdot a_{0i_n} \ ,\\
\cH_{Pt}(q_0) \es
\sum_{n=2}^\infty \, 
\sum_{i_1, i_2, ..., i_n} \, c_t(i_1,...,i_n) \, 
\left( {1 \over
2}\sum_{k=1}^n p_{i_k}^+ \right)^2 \, \, q^\dagger_{0i_1}
\cdot \cdot \cdot q_{0i_n} \ .
\end{eqnarray}
The operator $\cH_f$ is called 
the free Hamiltonian, since it is the part of $\cH_0(a_0)$ that does not depend on the coupling constant. The index $i$ denotes the quantum numbers of the particle 
and $p_i^-$ is the free FF energy for the gluon 
kinematical momentum components $p_i^+$ and 
$p_i^\perp$.

In order to solve this equation perturbatively, the Hamiltonian can be expressed as an expansion in powers of the coupling constant $g$:
\begin{eqnarray}
\cH_t 
&=&
\cH_0 + g\cH_{t1} + g^2\cH_{t2} + g^3\cH_{t3} + g^4\cH_{t4} + \dots \ 
\end{eqnarray}
In an expansion up to fourth-order, the RGPEP equation reads,
\begin{eqnarray}
&& \cH_{0}' + g\cH_{t1} + g^2\cH_{t2}' + g^3\cH_{t3}' + g^4\cH_{t4}' \nn
&=&\left[\left[\cH_0,\cH_0 + g\cH_{1P t} + g^2\cH_{2P t} + g^3\cH_{3P t} + g^4\cH_{4P t} \right],\cH_0 + g\cH_{t1} + g^2\cH_{t2} + g^3\cH_{t3} + g^4\cH_{t4} \right] \ .
\end{eqnarray}
Collecting terms with the same power of $g$ one can solve the equation order by order, starting from order zero and introducing the result in the next-order equation:
\begin{eqnarray}
\cH_0' \es 0 \ , \\
g\cH_{t\,1}' 
\es \left[\left[ \cH_0, g\cH_{1P t}\right],\cH_0\right] \ , \\
g^2\cH_{t\,2}' 
\es \left[\left[ \cH_0, g^2\cH_{2P t}\right], \cH_0\right] + \left[\left[ \cH_0, g\cH_{1P t}\right],g \cH_{1 t}\right] \ , \\
g^3 \cH_{t\,3}' 
\es \left[\left[ \cH_0, g^3\cH_{3P t}\right], \cH_0\right] + \left[\left[ \cH_0, g^2\cH_{2P t}\right], g\cH_{1 t}\right] +
\left[\left[ \cH_0, g\cH_{1P t}\right], g^2\cH_{2 t}\right]  \ ,\\
g^4 \cH_{t\,4}' \es \left[\left[ \cH_0, g^4\cH_{4P t}\right], \cH_0\right] + \left[\left[ \cH_0, g^3\cH_{3P t}\right], g\cH_{1 t}\right] +
\left[\left[ \cH_0, g^2\cH_{2P t}\right], g^2\cH_{2 t}\right] + \left[\left[ \cH_0, g\cH_{1P t}\right], g^3\cH_{3 t}\right] \ ,
\end{eqnarray}
The solutions to these equations in terms of matrix elements $\cH_{t\,ab}=\langle a|\cH|b\rangle$ are:
\begin{eqnarray}
\cH_{t1\,ab} 
\es
f_{t\,ab}\,\cH_{01\,ab}  \ ,
\\
\cH_{t2\,ab} 
\es  f_{t\,ab}\, \sum_x
\cH_{01\,ax}\cH_{01\,xb}\, \cB_{t\,axb}
\ + \ f_{t\,ab}\,\cG_{02\,ab}  \ ,
\\
\cH_{t3\,ab} 
\es 
f_{t\,ab}\,\cG_{03\,ab}  +  
f_{t\,ab}\,
\sum_{xy}\cH_{01\,ax}\,\cH_{01\,xy}\,\cH_{01\,yb} \, \cC_{t\,axyb}
\, + \,
f_{t\,ab}\,\sum_{x} 
\,
\left(
\cH_{01\,ax}\,\cG_{02\,xb}
 +  \cG_{02\,ax}\,\cH_{01\,xb}
\right)\,  \cB_{t\,axb}  \ , \\
\cH_{t4\,ab} 
\es 
f_{t\,ab}\,\sum_{xyz} \cH_{01\,ax}\,
\cH_{01\,xy}\,\cH_{01\,yz}\,\cH_{01\,zb} \,\,\cD_{t\,axyzb}
\np
f_{\tau\,ab}\,\sum_{xy} \left(\cH_{01\,ax}\,
\cH_{01\,xy}\,\cG_{02\,yb}
\ + \ \cH_{01\,ax}\,\cG_{02\,xy}\,\cH_{01\,yb}\ + \ \cG_{02\,ax} \,\cH_{01\,xy}\cH_{01\,yb}
\right)
\, \cC_{t\, axyb}
\np 
f_{\tau\,ab}\,\sum_x  \left(\cG_{02\,ax} 
\,\cG_{02\,xb}+ \cH_{01\,ax}\,\cG_{03\,xb} + \cG_{03\,ax}\, \cH_{01\,xb}\right)
\, \cB_{t\, axb}
\ + \ f_{\tau\,ab}\, \cG_{04\,ab} \ . 
\end{eqnarray}
The RGPEP factors $\cA_{t\,axb}$, $\cB_{t\,axb}$, $\cC_{t\,axyb}$ and $\cD_{t\,axyzb}$ result from integration and are given below. They containing in turn form factors  $f_{t\, ab}:=\exp\left[-ab^2\, t\right]$, with the symbol $ab$ denoting differences of invariant masses squared, $ab:=\cM_{ab}^2-\cM_{ba}^2$. Every $\cG_{t\,ab}$ is defined through $\cH_{t\,ab}=f_{t\,ab}\cG_{t\,ab}$ .
\begin{eqnarray}
\cA_{t\,axb} 
\es
\left[ax\, p_{ax}\, + bx\, p_{bx}\right] f_{t\,ab}^{-1}f_{t\,ax}f_{t\,bx} \ , 
\\
\cB_{t\,axb} 
\es  
\int_0^t \cA_{\tau\,axb}\, d\tau \ , 
\\
\cC_{t\,axyb} 
\es 
\int_0^t\left[\cA_{\tau\,axb} 
  \,\cB_{\tau\,xyb} \ + \ 
\cA_{\tau\,ayb}\,\cB_{\tau\,axy}\right] d\tau \ , \\
\cD_{t\,axyzb} \ ,
\es
\int_0^t \left[\cA_{\tau\,axb}\, \cC_{\tau\,xyzb} 
\ + \
\cA_{\tau\,azb}\,
 \cC_{\tau\,axyz} 
\ + \ 
A_{\tau\,ayb}\,
\, \cB_{\tau\,axy} \,
\,\cB_{\tau\,yzb} \right] \ .
\end{eqnarray}
The alphabetical indices denote configurations of particles, i.e. a collection of quantum numbers that label creation and annihilation operators. For more details about the notation see Ref.~\citep{GlazekPerturbative}.

\end{appendix}

\end{document}